\documentclass[prc,showpacs,twocolumn,aps]{revtex4}

\usepackage{bm}
\usepackage{amsmath}
\usepackage{epsfig}
\usepackage{psfig}
\usepackage{graphics}
\usepackage{graphicx}

\begin{document}

 ~ \\  \begin{flushright}TRIUMF preprint: TRI-PP-04-26\end{flushright}

\title{Nucleon-nucleus optical potential in the particle-hole approach} 

\author{C.~Barbieri}
  \email{barbieri@triumf.ca}
  \affiliation
     {TRIUMF, 4004 Wesbrook Mall, Vancouver,
          British Columbia, Canada V6T 2A3 \\  }

\author{B.~K.~Jennings}
  \affiliation
     {TRIUMF, 4004 Wesbrook Mall, Vancouver,
          British Columbia, Canada V6T 2A3 \\  }

\date{\today}

\begin{abstract}
 Feshbach's projection formalism in the particle-hole model space leads to
a microscopic description of scattering in terms of the many-body
self-energy.
To investigate the feasibility of this approach,
an optical potential for ${}^{16}$O is constructed starting
from two previous calculations of the self-energy
for this nucleus.
The results reproduce the background phase shifts for
positive parity waves and the resonances beyond the mean field.
 The latter can be computed microscopically for
energies of astrophysical interest using Green's function theory.
\end{abstract}

\preprint{TRI-PP-04-26}

\maketitle

\section{Introduction}
\label{sec:intro}

 The Feshbach's theory of the optical potential~\cite{fesh} provides
a tool to describe the scattering of nucleons from nuclei.
 In its original form, the full Hilbert space is partitioned
in a subspace that contains only one particle added to a particular
state of the core nucleus.  This includes the elastic scattering states and the
nuclear orbits unoccupied by the target's nucleons.
 The resulting optical potential can be thought as an effective interaction
that accounts for the effects of the degrees of freedom of the excluded
space
---such as the overall antisymmetrization of the wave function, excitations
of the target or breakup channels. 
The complexity of these effects makes the {\em ab initio} calculation of the
optical potential a very difficult task.
The choice of working in the space of one particle plus a
core is shared by several theories of nucleon-nucleus scattering.
Examples are the cluster model~\cite{cluster1},
folding potential~\cite{AdvNP_kara} and shell model embedded
in the continuum~\cite{SMEC_PhysRep}.
All these techniques have been successfully applied to scattering processes.

A conceptually different approach consists in extending the scattering space to
include both a particle on top of the nuclear core and
the possibility of propagating a hole excitation. 
In the following, we will refer to this as the particle-hole ({\em ph})
Hilbert space.
Mahaux and Capuzzi~\cite{MaCa} and Jennings and Escher~\cite{byron02} have 
shown that applying the Feshbach's formalism to this space leads to
an optical potential that is the usual many-body self-energy defined
in Green's function theory~\cite{AAA,fetwa}.
The properties of using the self-energy as an optical potential have been
discussed by Mahaux and Sartor in Ref.~\cite{Mahaux}.
The solutions of the scattering equation above (below) the Fermi energy are
the overlap wave functions between the core and the eigenfunctions of the
systems with $A+1$ ($A-1$) nucleons. For particle states these are the same
wave functions as obtained in the original Feshbach approach~\cite{fesh}. 
The fact that states with $A-1$ particle are included in the formalism does not
lead to complications in the study of scattering events.
 Rather, more physical information can be extracted using the self-energy
since it also describes the so called `Pauli forbidden' orbitals,
occupied by the nucleons of the target.
Correspondingly, the optical potential in the {\em ph} space is more easily
comparable to phenomenological models (based on, e.g.,
Wood-Saxon wells)~\cite{Mahaux,MaCa} since these also describe hole states.
It also has better analytical properties than its counterpart in the
particle-only space~\cite{MaCa,byron02}.
Finally, the theory of Green's functions provides a natural way to include
the effects of the excitations of the core in terms of an expansion
in Feynman diagrams~\cite{Mahaux,diba04}.

Many-body Green's functions have been applied in the past to
study nuclear correlations, with emphasis on the hole part of the
one-body spectral function.
 Recent developments offer the opportunity to obtain sophisticate
descriptions of the couplings between particles and
collective states using a Faddeev expansion~\cite{BaDi01,BaDi03}.
 A similar formalism was already considered in Ref.~\cite{Dan94}
for the optical potential where, however, the dressing of propagators
was disregarded and no application was attempted.
 The calculations of Refs.~\cite{BaDi01,BaDi02} account for collective motion
near the Fermi level, including the energy regime of interest to nuclear
astrophysics.
 With a variety of exotic isotopes that are involved in stellar processes
becoming experimentally accessible in modern radioactive beam facilities,
it is important to investigate whether the Green's function approach
can be applied to study low-energy nucleon scattering and capture processes.

We have recently considered the self-energy resulting from a recent application
of the self-consistent Green's function (SCGF) method
to ${}^{16}$O~\cite{BaDi02} and explored its predictions
for proton-nucleus scattering.
 These results were obtained in a restricted model space,
which is not fully appropriate to describe scattering events.
However, preliminary calculations with only this input gave encouraging
results~\cite{BaJeNIC8}.
In this work the self-energy of Ref.~\cite{BaDi02} is augmented by including
the components from outside this model space, as they have been computed
in Ref.~\cite{DMP}. We then report on the final conclusions of these
exploratory studies.

The model and the details of the calculations are given in
Sec.~\ref{sec:model}.
The results for elastic scattering and the bound states of $^{17}$F
are reported in Sec.~\ref{sec:results} and are preceded by a discussion
of the different contributions to the self-energy, in Sec.~\ref{sec:modeldpns}.
A discussion and our conclusion are given in Secs.~\ref{sec:disc}
and~\ref{sec:concl}.

\section{Model}
\label{sec:model}

\begin{figure}[t]
 \begin{center}
 \includegraphics[width=8.4cm]{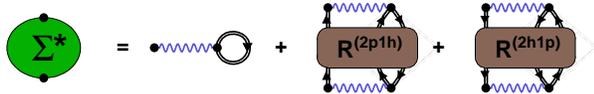}
 \end{center}
 \vspace{-.3in}
\caption[]{
  Feynman diagrams representation of the self energy. The first diagram 
 on the r.h.s. represents the Hartree-Fock like contribution to the mean field.
 The remaining ones describe core polarization effects in the particle (2p1h)
 and hole (2h1p) part of the spectrum.}
\label{fig:slef-en}
\end{figure}

\begin{figure}[ht]
 \begin{center}
\includegraphics[height=6.5cm,width=4.5cm]{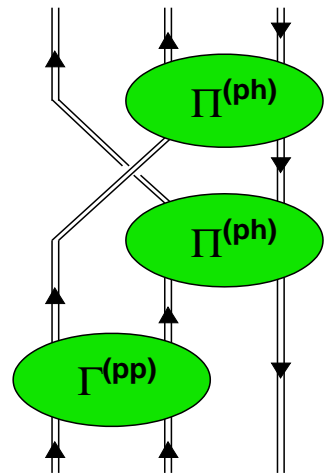}
 \end{center}
\caption[]{
  Example of a diagrammatic contribution included in the Faddeev expansion
  for $R^{(2p1h)}$ (see Fig.~\ref{fig:slef-en}). A quasiparticle is coupled
  to the response function $\Pi^{(ph)}$ that describes the target nucleus.
  It can also participate in a pairing processes,
  which is accounted by the two-body propagator~$g^{II,(pp)}$.}
\label{fig:FaddEx}
\end{figure}

In the SCGF approach it is useful to split the many-body self-energy into
three contributions~\cite{winter,diba04}, as shown diagrammatically
in Fig.~\ref{fig:slef-en}.
There, the double lines represent the exact one-body Green's function,
which contains complete information on the particle and hole spectral
distributions.
 The first diagram on the r.h.s. is the direct extension of the Hartree-Fock
potential to include the effects of the fragmentation of strength and
represents the nuclear mean-field (MF) in the presence of correlations.
The remaining contributions split naturally in diagrams containing
{\em at  least} two-particle--one-hole (2p1h), describing the system
of A+1 particles, or two-hole--one-particle (2h1p), corresponding to
A-1 particles.
 The irreducible propagators $R^{(2p1h)}$ and $R^{(2h1p)}$  account for
the core polarization contributions to the optical potential in the particle
and hole spaces, respectively~\cite{Mahaux}.
 The separation of Fig.~\ref{fig:slef-en} is exact. 
 In Refs.~\cite{BaDi01,BaDi02}, $R^{(2p1h)}$ and $R^{(2h1p)}$ were
computed employing a Faddeev expansion that permits the direct coupling of the
single-particle motion to collective excitations of the core. These were
evaluated in the dressed random phase approximation (DRPA)~\cite{DRPA}.
 The example of a diagram that contributes to $R^{(2p1h)}$ is given in
Fig.~\ref{fig:FaddEx}.
 Since this expansion is based on the fully fragmented single particle
propagator ---which is generated from the self-energy itself---
a self-consistent solution is required.

The nuclear self-energy computed in Ref.~\cite{BaDi02} was obtained
within a model space $\cal{P}$ corresponding to the  harmonic oscillator
wave functions for all orbitals up to the $pf$ shell plus the $g_{9/2}$.
A parameter $b$=1.76~fm was employed.
 This space appears to be large enough to describe the influence of the
low energy (long-range) excitations on nuclear fragmentation~\cite{BaDi03}.
However, it requires a proper extension for applications to single particle
scattering, as it will be discussed below.
The effect of correlations outside this model space were accounted for
by employing a G-matrix as an effective interaction, which was derived from
the Bonn-C potential~\cite{bonnc} according to Ref.~\cite{CALGM}.
The computation of the G-matrix for positive energies is an outstanding
problem which was not attempted there.
 Therefore, we employed a fixed starting energy of -5MeV in the present work,
as the closest reliable choice to the continuum.

At low energies the optical potential is well approximated by a real
interaction and $R^{(2p1h)}$ and $R^{(2h1p)}$  can be expressed as discrete
sums of poles. Thus, for each given partial wave, $lj$, the contributions
depicted in Fig.~\ref{fig:slef-en} can be expressed as
\begin{subequations}
\label{eq:Self-en_Fadd}
\begin{equation}
 \Sigma^{MF,Fadd}_{lj}(k,k') =
 \sum_{n_\alpha, n_\beta \in \cal{P}} \; \phi_\alpha(k) \; 
         \Sigma^{MF,Fadd}_{lj;n_\alpha, n_\beta} \; 
     \phi^*_\beta(k')  \; \; ,
\label{eq:Self-en_Fadd_1}
\end{equation}
\begin{equation}
 \Sigma^{(2p1h),Fadd}_{lj}(k,k') =
 \sum_{n_\alpha, n_\beta \in \cal{P}}  \phi_\alpha(k)
       \left[
        \sum_{n+} \frac{ \left( m^{n+}_\alpha \right)^* \; m^{n+}_\beta}
                      {\omega - \varepsilon^{n+}_{lj} + i\eta}
       \right]
     \phi^*_\beta(k')  \; \; ,
\label{eq:Self-en_Fadd_2}
\end{equation}
\begin{equation}
 \Sigma^{(2h1p),Fadd}_{lj}(k,k') =
 \sum_{n_\alpha, n_\beta \in \cal{P}}  \phi_\alpha(k)
       \left[
        \sum_{k-} \frac{ \left( m^{k-}_\alpha \right)^* \; m^{k-}_\beta}
                      {\omega - \varepsilon^{k-}_{lj} - i\eta}
       \right]
     \phi^*_\beta(k')  \; \; ,
\label{eq:Self-en_Fadd_3}
\end{equation}
\end{subequations}
where $\phi_\alpha(r)$ are the harmonic oscillator radial functions
referring to single particle quantum numbers
\hbox{$\alpha=\{n_\alpha, l_\alpha, j_\alpha, m_\alpha\}$}~%
\footnote{The isospin degrees of freedom are not shown explicitly here.},
the first sum runs over all the orbits belonging to the model space and
$l_\alpha j_\alpha=l_\beta j_\beta=lj$ since ${}^{16}$O has a $0^+$
isoscalar ground state.

The superscript ``{\em Fadd}'' indicates that Eqs.~(\ref{eq:Self-en_Fadd})
represent the results of Ref.~\cite{BaDi02}. This is the most sophisticated
calculation available to date for the self-energy at low-energies
that account for the coupling between single nucleons and collective
excitations.
 However, the expansion over a few harmonic oscillator states is not optimal
for describing the details of the nuclear surface. Analogously, it misses
part of the large momentum components in the optical potential.
This is particularly critical for the MF component%
, which describes the background of the  phase shifts.
On the contrary, the
same nucleus was studied in Refs.~\cite{DMP} employing a spherical box basis
that includes all the relevant momentum components. An effective G-matrix,
derived for nuclear matter and the Bonn-B potential~\cite{bonnc}, accounted
for the binding due to short-range and tensor correlations.
 The self-energy, computed only to second order in the
perturbation series, neglected most of the collective effects.
 This approach was applied to obtain the quasihole wave functions
associated to the $p$ states occupied in ${}^{16}$O, with 
sufficiently accurate results to describe the shapes of the ($e,e'p$)
cross sections to those states~\cite{RadiciWim}.

In this work, we chose to employ a mixed representation of the self-energy
in which the MF components missing in the space $\cal{P}$ were extracted
from Refs.~\cite{DMP}, while the contributions beyond MF computed in
Ref.~\cite{BaDi02} [Eqs.~(\ref{eq:Self-en_Fadd_2})
and~(\ref{eq:Self-en_Fadd_3})] were retained.
 To do this the MF self-energy of Ref.~\cite{DMP} was split in two parts,
\begin{equation}
  \Sigma^{MF , Box}_{lj}(k,k';\omega) =
     \Sigma^{MF,Box}_{0,lj}( k,k';\omega) 
   + \Sigma^{MF,Box}_{1,lj}( k,k';\omega) \; ,
\label{eq:DMP_MF}
\end{equation}
where $\Sigma^{MF,Box}_{0}$ is the projection onto  ${\cal P}$
and $\Sigma^{MF,Box}_{1}$ acts on the excluded space.
Two approximations were considered depending on which MF component to employ
inside ${\cal P}$.
In the first case (I), $\Sigma^{MF,Box}_{1}$ was added to
Eq.~(\ref{eq:Self-en_Fadd_1}).
In doing this, we note that the G-matrix used to compute $\Sigma^{MF,Fadd}$
accounts for the extra binding due to the degrees of freedom of the excluded
space.
 Since these are reinserted explicitly by $\Sigma^{MF,Box}_{1}$, one should
also rescale $\Sigma^{MF,Fadd}$ appropriately by a constant,~$N^I$.
The second choice (II), consisted in employing both parts of
Eq.~(\ref{eq:DMP_MF}). Also in this case we kept the possibility of tuning 
the depth of the potential.
The complete MF  contributions employed in this work are
\begin{subequations}
\label{eq:Self-en_MF}
\begin{equation}
 \Sigma^{MF,I}_{lj}(k,k';\omega) =
             N^I_{lj} \;  \Sigma^{MF,Fadd}_{lj}(k,k')  ~+~
            \Sigma^{MF,Box}_{1,lj}(k,k')  \;  , 
\label{eq:Self-en_MF_I}
\end{equation}
\begin{equation}
  \Sigma^{MF,II}_{lj}(k,k';\omega) =
             N^{II}_{lj} \; \Sigma^{MF,Box}_{lj}(k,k')  \; ,
\label{eq:Self-en_MF_II}
\end{equation}
\end{subequations}
where the constants $N^I_{lj}$ and $N^{II}_{lj}$ depend of the specific
channel and will be discussed below.
The full self-energy employed in the calculations is [see
Fig.~\ref{fig:slef-en}]
\begin{eqnarray}
\lefteqn{
  \Sigma^{\star,I(II)}_{lj}(k,k';\omega) ~=~ \Sigma^{MF,I(II)}_{lj}(k,k')
        } 
\nonumber \\
 & & + \Sigma^{(2p1h),Fadd}_{lj}( k,k';\omega)    +
  \Sigma^{(2h1p),Fadd}_{lj}( k,k';\omega)   \; .
\label{eq:Self-en_Full}
\end{eqnarray}

The Dyson equation can be expressed in a Schr\"odinger-like form, where
the self-energy takes the place of a non-local and energy dependent optical
potential~[$\hbar=c=1$ and $\mu$ is the reduced mass]
\begin{widetext}
\begin{equation}
  \frac{k^2}{2 \mu}  \; \psi(k)
  ~+~
  \int_0^\infty dk' \; k'^2
  \left\{
       \; \Sigma_{lj}^\star(k,k';E_{cm}) 
     \; + \; V^l_{Coul.}(k,k') 
  \right\} \; \psi(k')
  ~=~ E_{cm} \; \psi(k)   \; ,
\label{eq:Dyson}
\end{equation}
\end{widetext}
where $V^l_{Coul.}(k,k')$ in the Coulomb interaction corresponding to a
uniformly charged sphere of radius $R_c$~=~3.1~fm.
This was added to account for the electromagnetic interaction missing
in the calculations of Refs.~\cite{BaDi02,DMP}.
Due to the non local character of $\Sigma^{\star}$, Eq.~(\ref{eq:Dyson}) is
conveniently solved in momentum space. In doing this, the long distance
part of the Coulomb potential was solved using the Kwon-Tabakin-Lande~\cite{KTL}
procedure for bound states and the Vincent-Phatak~\cite{VP} one for scattering.

Above the Fermi level the eigenvalues of Eq.~(\ref{eq:Dyson}) are
related to the spectrum of ${}^{17}$F
by $E^n_{cm} =E^{{}^{17}F}_{n} - E^{{}^{16}O}_{g.s.}$.
Thus, $E_{cm} > 0$ describes the scattering of protons from ${}^{16}$O
while the bound solutions are the overlaps of the ground state of ${}^{16}$O
with the corresponding bound states ${}^{17}$F.
Analogously, below the Fermi level
$E^n_{cm} =E^{{}^{16}O}_{g.s.} - E^{{}^{15}N}_{n}$ and the eigenstates
represent the overlaps with ${}^{15}$N.
The Dyson equation implies that the bound solutions of Eq.~(\ref{eq:Dyson})
have to be normalized to their spectroscopic factor according to
\begin{equation}
  Z_{lj}^n = \int_0^\infty dk \; k^2 \left| \psi^n(k) \right|^2 =
   \left[ 1 - 
     \left.
     \left\langle \tilde{\psi}^n \right| \frac{d \Sigma_{lj}^\star}{d\omega}
     \left|       \tilde{\psi}^n \right\rangle
     \right|_{\omega=E_{cm}^n}
     \right]^{-1}  \; ,
 \label{eq:Spect_fac}
\end{equation}
where $\tilde{\psi}^n(k)$ is the solution itself normalized to unity
and $E_{cm}^n$ is the corresponding eigenvalue.
The asympotic normalization for the unbound solutions is related in the usual
way to the flux of incoming particles.

\section{Results}
\label{sec:results}

Eqs.~(\ref{eq:Self-en_MF}) and~(\ref{eq:Self-en_Full}) include the relevant
physics from both the calculations of Refs.~\cite{BaDi02} and~\cite{DMP}.
This self-energy represents a model for the optical potential
that acts on the full {\em ph} Hilbert space and can give sensible predictions
near the Fermi level.
However, the two-body realistic interactions alone, as used in these works, cannot reproduce
the experimental binding energies and spin-orbit splitting for nuclei 
with $A\ge3$~\cite{GFMC,ShellMod}.
To obtain these, relativistic effects or three-body forces are
required~\cite{MutVj}.
In this work $\Sigma^{\star}$ was constrained to reproduce the experimental
spectrum in two ways.
First, the constants $N^I_{lj}$ and $N^{II}_{lj}$ that affect the depth
of the optical potential were chosen to reproduce the corresponding
quasiparticle energies.
These are the $s_{1/2}$ and $d_{5/2}$ bound states of ${}^{17}$F,
its $d_{3/2}$ resonance and the $p_{1/2}$ and  $p_{3/2}$ hole states of
${}^{15}$N.
Second, complex resonances that do not have a mean field
character are generated by the dynamic part of the self-energy.
At low energy, most of these couple to only one pole $\varepsilon^{i\pm}$
in Eqs.~(\ref{eq:Self-en_Fadd_2}) and~(\ref{eq:Self-en_Fadd_3}).
 Therefore, we have fitted those poles that could be identified with specific
resonances of the A+1 system (${}^{17}$F) by imposing that
Eq.~(\ref{eq:Dyson}) yields the corresponding experimental energies.
%
%
%
We note that a similar approach was already employed in
Ref.~\cite{BaDi02}. This is necessary for the particular case of ${}^{16}$O
due to the strong coupling between the single particle spectrum and
collective motions, which suggest the need for an improved description 
of the low-energy structure of this nucleus~\cite{BaDi03} and more
attractive effective interactions~\cite{Martino02}.
 Although, satisfactory results can already be obtained in similar
calculations for heavier nuclei~\cite{Brand,Jie,Sn132}.

\begin{table}
\begin{ruledtabular}
\begin{tabular}{lccccccccc}
   &  & $\Sigma^{MF,Fadd}$ & &  
    $\Sigma^{MF,Fadd}$   & &
    $\Sigma^{\star,I}$   & &  &   \\
 $ lj    $ &  & $E^{lj}_{cm}$~(MeV)\footnotemark[1] & &  
    $N^I_{lj}$\footnotemark[2]  
      & &
    $N^I_{lj}$\footnotemark[2]   & &
    $E^{exp}_{cm}~(MeV)$  &   \\
  \hline  
 $s_{1/2}$ &  &    -3.57 & &  0.69  & & 1.05 & &   -0.1   &   \\
  \\
 $d_{3/2}$ &  &     1.87 & &  0.72  & & 1.08 & &    4.4   &   \\
 \\
 $p_{3/2}$ &  &   -16.61 & &  1.06  & & 1.07 & &   -18.5  &   \\
           &  &          & &        & & 0.95 ($E^{p_{3/2}}_{cm}$=-15.1) &     &   \\
\end{tabular}
\end{ruledtabular}
\footnotetext[1]{$N^I_{lj} = 1$,}
\footnotetext[2]{$E^{lj}_{cm}\equiv E^{exp}_{cm}$, except when specified.}
 \caption[]
  {Corrections applied to the depth of the MF potential $\Sigma^{MF,I}$
  [Eq.~(\ref{eq:Self-en_MF_I})] and quasi particle energies obtained
  in the calculations of Fig.~\ref{fig:MFdep_mix}. }
 \label{tab:MFdep}
\end{table}

The influence of this fitting procedure on the results is discussed in the
following.
After calibrating Eqs.~(\ref{eq:Self-en_Fadd}) and~(\ref{eq:Self-en_MF})
to the spectra of ${}^{17}$F and ${}^{15}$N, 
the results for the scattering phase shifts and the bound single particle
wave functions are a prediction of the model.

\begin{figure}[t]
 \includegraphics[width=8.cm]{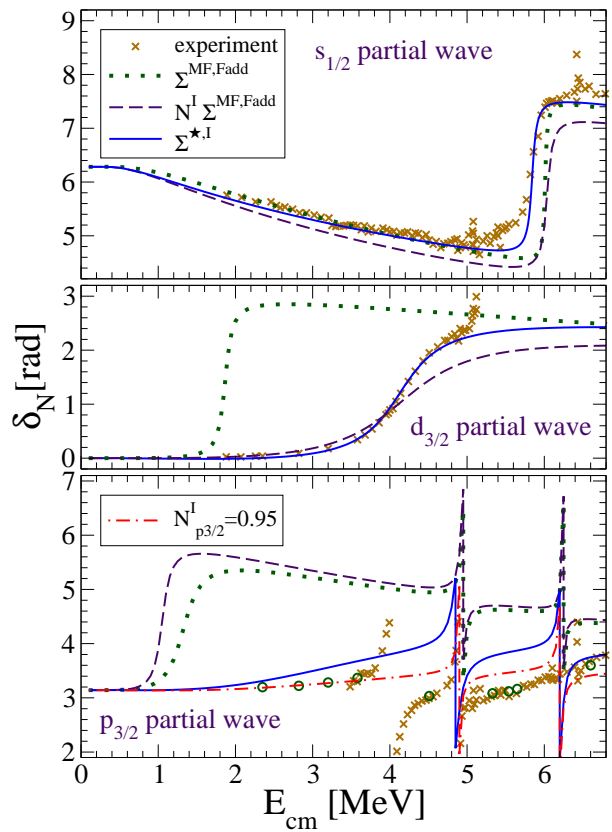}
 \caption{Phase shifts obtained from the self consistent self-energy of
 Ref.~\cite{BaDi02} [Eq.~\ref{eq:Self-en_Fadd}] before (dotted lines)
 and after (dashed lines) correcting the depth of $\Sigma^{MF,Fadd}$
 to reproduce the quasiparticle (and quasihole) energies.
  The full lines are obtained when also the momentum components outside
 the model space ${\cal P}$ are included, Eq.~(\ref{eq:Self-en_MF_I}).
  The $s_{1/2}$, $d_{3/2}$ and  $p_{3/2}$ partial waves are shown. For 
 $p_{3/2}$, the dot-dashed line was obtained by fitting $N^I_{p_{3/2}}$ to
 reproduce the background phase shifts rater than the quasihole energy.
  The values of the corrections $N^I_{lj}$ are reported in
  Tab.~\ref{tab:MFdep}.
  The experimental results are from Refs.~\cite{Salisb62} (crosses)
 and~\cite{Blue} (circles). }
 \label{fig:MFdep_mix} 
\end{figure}

\subsection{Parameter dependence}
\label{sec:modeldpns}

To discuss the influence of the different contributions to 
Eq.~(\ref{eq:Self-en_MF}), 
the phase shifts for proton scattering have been computed employing different
truncations of the mean field self-energy $\Sigma^{MF,I}$.
 The results are shown in Fig.~\ref{fig:MFdep_mix} for three partial waves.
The dotted lines were obtained by retaining only the original contribution
to the self-energy of Ref.~\cite{BaDi02}. 
Thus, neglecting $\Sigma^{MF,Box}_1$ in Eq.~(\ref{eq:Self-en_MF_I}) and
setting $N^I_{lj}=1$ for all cases.
The results obtained by constraining these constants to generate the
proper quasiparticle energies is given by the dashed lines.
The full line shows the full results form Eq.~(\ref{eq:Self-en_MF_I}), obtained
by including also the $\Sigma^{MF,Box}_1$ term and refitting the $N^I_{lj}$.
The values for the quasiparticle energies and the constants $N^I_{lj}$ used
are given in Table~\ref{tab:MFdep}

The background contribution to the phase shifts of the $s_{1/2}$ partial wave
is described correctly by $\Sigma^{MF,Fadd}$ but not the energy of the bound
state. Vice versa, it is possible to constrain the depth of the potential
to reproduce the latter but the agreement with the experimental phase shifts
is lost.  However, both quantities are reproduced if $\Sigma^{MF,Box}_1$ is included.
 In this case the correction required in the depth of the potential,
$N^I_{S_{1/2}}=1.05$, is less significant than when only $\Sigma^{MF,Fadd}$
is included.
A similar trend is seen for the $d_{3/2}$ channel. Reproducing the energy
of the  single particle resonance with  $\Sigma^{MF,Fadd}$ alone
requires a sizable change in its depth, while
the observed phase shifts are obtained only after including
the components outside the space~$\cal{P}$.
We observe that the expansion of Eq.~(\ref{eq:Self-en_Fadd}) includes only
one harmonic oscillator function for the $d_{3/2}$ wave and two for
$s_{1/2}$. With such a restricted space, it is remarkable that
the resulting background phase shifts are still obtained somewhat close to the
experiment.

A different behavior is found for the $l=1$ partial waves.  The results
for $p_{3/2}$ are shown in Fig.~\ref{fig:MFdep_mix}~(the $p_{1/2}$ case is analogous).
 In this cases $\Sigma^{MF,Fadd}$ produces a spurious resonance at
$\sim$1~MeV that is not seen experimentally. Fitting the potential's depth
to constrain the quasihole energies of ${}^{15}$N generates a more attractive
well, thus worsening the situation. The phase shifts improve upon
introducing $\Sigma^{MF,Box}_1$ (full line) but still show a rise of the
background with the cm energy, while the experimental results are practically
constant. A proper choice of $N^I_{p_{3/2}}$ (and $N^I_{p_{1/2}}$) allows to
reproduce the behavior of the phase shifts at the lower energies
but results in underbinding
the corresponding orbitals in ${}^{16}$O (see Tab.~\ref{tab:MFdep}).

 The curves of Fig.~\ref{fig:MFdep_mix} have been computed without any
shift of the $\varepsilon^{i+}$ poles in in Eq.~(\ref{eq:Self-en_Fadd_2}).
This gives an idea of the quality the energy spectra obtained 
adopting the interaction of Ref.~\cite{CALGM}.
No solutions were obtained that could be interpreted as the $d_{3/2}$
resonances above 5~MeV.
 The $s_{1/2}$ resonance at $\sim$6~MeV was obtained as a coupling
of a proton to the first excited state of ${}^{16}$O.
 Analogously, the two lowest resonances in both $p_{1/2}$ and $p_{3/2}$ can
be interpreted as quasiparticle interacting with the first isoscalar $3-$ and
$1-$ levels of ${}^{16}$O~\cite{BaF17-03}.

\subsection{Phase shifts for proton scattering}
\label{sec:results_phsh}

\begin{figure}[t]
 \includegraphics[width=8.cm]{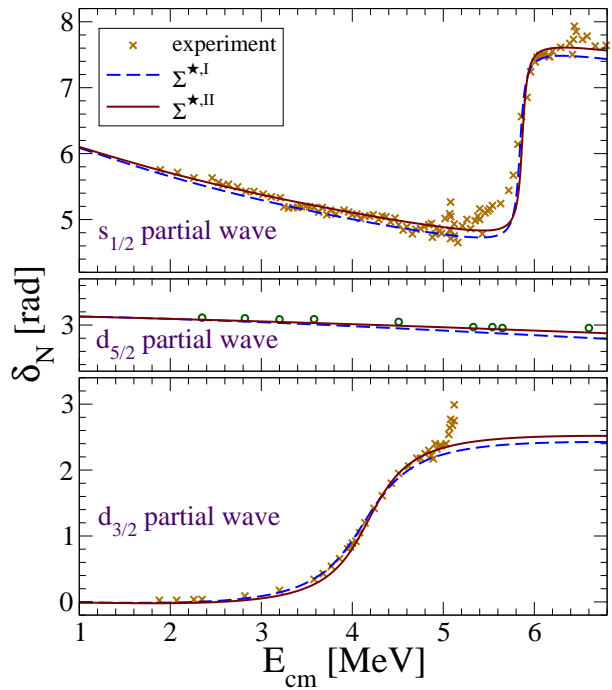}
 \caption{Phase shifts for positive parity waves obtained from the 
  the self-energies I~(dashed lines) and II~(full lines).
   The experimental results are from Refs.~\cite{Salisb62} (crosses)
   and~\cite{Blue} (circles). }
 \label{fig:phsh_final_+}
\end{figure}

\begin{figure}[t]
 \includegraphics[width=8.cm]{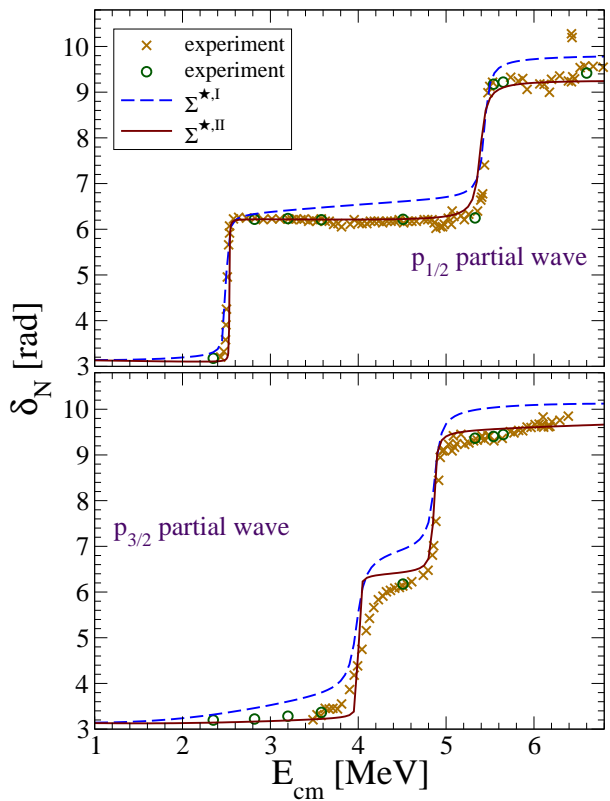}
 \caption{Phase shifts for negative parity waves obtained from the 
the self-energies I~(dashed lines) and II~(full lines). 
   The experimental results are from Refs.~\cite{Salisb62} (crosses)
   and~\cite{Blue} (circles). }
 \label{fig:phsh_final_-}
\end{figure}

 Figures~\ref{fig:phsh_final_+} and \ref{fig:phsh_final_-} compare the 
phase shifts obtained from both the Self-energies I and II after 
constraining the quasiparticle energies and resonances to
their experimental values. Table~\ref{tab:phsh_final} shows the values of 
the constants $N^I_{lj}$ and~$N^{II}_{lj}$ used to obtain these results.

For the positive parity waves the background phase shifts are described
equally well by both optical potentials.
 The potential $\Sigma^{\star,II}$ can also describe the negative parity
waves and it is more accurate for the $p_{1/2}$ case, for which the collective
resonances are sharper.
%
In general, the non MF resonances were predicted narrower than the
experiment. This is probably related to the lack of momentum components
outside the model space ${\cal P}$ in Eqs.~(\ref{eq:Self-en_Fadd_2})
and~(\ref{eq:Self-en_Fadd_3}), which were not corrected as for the MF part
of the self-energy.

The values of $N^I_{lj}$ and~$N^{II}_{lj}$ show that much smaller modifications
are needed to force $\Sigma^{\star,I}$ to reproduce the quasiparticle energies.
This is consistent with the more sophisticate treatment of long-range
correlations achieved in Ref.~\cite{BaDi02}.
 It is worth noting that the inability to describe both the bound energies
and the scattering for the negative parity waves is consistent with the lack
of three body forces in the present model, which are needed to reproduce
their spin-orbit splitting.
 On the other hand the results with the MF potential II reproduce
reasonably well both the quasihole energies and the phase shifts.

\begin{table}
\begin{ruledtabular}
\begin{tabular}{lccccccc}
   & &  
    $\Sigma^{\star,I}$   & &
    $\Sigma^{\star,II}$   & &  &   \\
 $ lj    $ &  & 
    $N^I_{lj}$\footnotemark[1]   & &
    $N^{II}_{lj}$\footnotemark[1]   & &
    $E^{exp}_{cm}$~(MeV)  &   \\
  \hline  
 $d_{3/2}$ &  &  1.08  & & 1.24 & &    4.4   &   \\
 \\
 $s_{1/2}$ &  &  1.05  & & 1.24 & &   -0.1   &   \\
  \\
 $d_{5/2}$ &  &  1.16  & & 1.29 & &   -0.6   &   \\
 \\
 $p_{1/2}$ &  &  1.06  & & 1.15 & &   -12.1  &   \\
 \\
 $p_{3/2}$ &  &  1.07  & & 1.25 & &   -18.5  &   \\
\end{tabular}
\end{ruledtabular}
\footnotetext[1]{$E^{lj}_{cm}\equiv E^{exp}_{cm}$, except when specified.}
 \caption[]
  {Corrections applied to the depth of the MF potentials I and II
  [Eq.~(\ref{eq:Self-en_MF})] for the calculation
  of Figs.~\ref{fig:phsh_final_+} and~\ref{fig:phsh_final_-}. }
 \label{tab:phsh_final}
\end{table}

\subsection{Bound overlap wave functions}
\label{sec:results_overl}

\begin{figure}
 \includegraphics[width=8.cm]{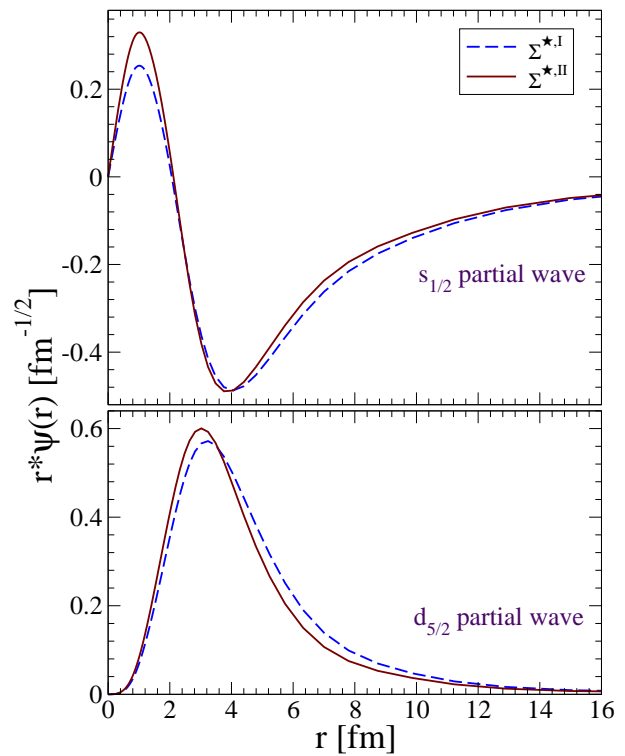}
 \caption{Radial part of the overlap wave functions between ${}^{16}$O and
 the bound $d_{5/2}$ and $s_{1/2}$ states of ${}^{17}$F.}
 \label{fig:bound_ovlp}
\end{figure}

The  overlap wave functions associated to the bound states
of ${}^{17}$F are shown in Fig.~\ref{fig:bound_ovlp} for the two choices
of Eq.~(\ref{eq:Self-en_MF}).
 The asymptotic behavior in presence of a Coulomb field is given by
\begin{equation}
  \psi_{lj}(r) \longrightarrow_{r \rightarrow \infty} 
       C_{lj}\frac{W_{-\eta, l+1/2}(r)}{r} \; ,
\label{eq:ANC_def}
\end{equation}
where $W_{-\eta, l+1/2}$ is a Whittaker function, $\eta$ the Sommerfield 
parameter and $C_{lj}$ the asymptotic normalization constant (ANC).
 The spectroscopic factors, ANCs and root-mean-square radii
obtained are given Tab.~\ref{tab:bound_ovlp}.

\begin{table}
\begin{ruledtabular}
\begin{tabular}{lcccccccc}
   & & \multicolumn{3}{c}{$s_{1/2}$} & 
     & \multicolumn{3}{c}{$d_{5/2}$} \\
   & &  $Z_{s_{1/2}}$ & $C_{s_{1/2}}$ & $\langle r_{s_{1/2}}^2 \rangle^{1/2}$ & 
     &  $Z_{d_{5/2}}$ & $C_{d_{5/2}}$ & $\langle r_{d_{5/2}}^2 \rangle^{1/2}$ \\
  \hline  
 $\Sigma^{\star,I}$ &  &  0.931  & -82.5 & 5.86 & & 0.913 & 1.07 & 4.01 \\
 \\
 $\Sigma^{\star,II}$ &  & 0.921  & -73.9 & 5.55 & & 0.909 & 0.81 & 3.70 \\
\end{tabular}
\end{ruledtabular}
 \caption[]
  {Spectroscopic factors,  ANCs (in fm$^{-1/2}$) and root-mean-square radii
 (in fm) for the bound $d_{5/2}$ and $s_{1/2}$ orbitals of ${}^{17}$F. }
 \label{tab:bound_ovlp}
\end{table}

 The Self-energy $\Sigma^{\star,I}$ predicts larger radii and ANCs
than $\Sigma^{\star,II}$,  which pulls these orbitals more strongly inside
the nucleus.  At the same time both choices yield the same spectroscopic factors,
implying equal occupancies.
 The depletion of these orbits is driven by the coupling to long-range
collective excitations contained in $\Sigma^{(2p1h),Fadd}$ and
$\Sigma^{(2h1p),Fadd}$. An additional quenching is expected from
short-range and tensor correlations and was not accounted for in this work.
%
 This has been seen to be of about 10\% for bound orbitals of several closed
shell nuclei~\cite{diba04}. 
However, what the strength of this reduction should be for
loosely bound nucleons, which can be largely localized at radii outside
the nuclear surface, has not yet been investigated.

\section{Discussion}
\label{sec:disc}

In order to include  the principals physics ingredients,
an improved self-energy was constructed from those obtained by
two different calculations of the nucleus of ${}^{16}$O. 
The effects of coupling a nucleon to collective modes were studied in
Ref.~\cite{BaDi02} but in a restricted model space.
 The  momentum components of the mean field potential outside this space
were instead extracted from Ref.~\cite{DMP}.
Since these calculations are based on realistic two-body
inter-nucleon reactions, the energy spectra cannot be accurately reproduced.
Therefore, the model has to be constrained phenomenologically to reproduce
the experimental spectra of the nuclei with $A\pm1$ nucleons.
 Resolving this situation may require the use of multi-nucleon forces and
more appropriate effective interactions.

 The present results show that both the inclusion of all momentum
components of the particle-hole Hilbert space and a proper treatment
of long-range correlations are important to correctly reproduce the 
mean field optical potential.
The coupling of single particle strength to long-range excitations is
also responsible for the creation of non mean field resonances.
After constraining the prediction for the single particle energies, the phase
shifts for the scattering of protons from ${}^{16}$O  were obtained in fair
agreement with the experimental data, except for the background behavior
of the of the $p$ waves.
 The difficulties for these waves are accompanied by the issue
of explaining the  hole spectroscopic factors with the same parity
extracted from $(e,e'p)$ experiments~\cite{leus}. 
  In Ref.~\cite{BaDi02} the latter were linked to the particularly complicate structure of the low-energy spectrum of ${}^{16}$O and further studies along this line have been initiated in Ref.~\cite{BaDi03}.
 It is plausible that the required improvements will resolve both the problems
of spectroscopic factors and scattering  phase shifts. 
Similar issues are expected to be beyond the requirements for
reproducing most heavier closed shell nuclei~\cite{Brand,rijs92}.

\section{conclusions}
\label{sec:concl}

 This work investigates the possibility of describing nucleon-nucleus
scattering  employing the many-body self-energy as an optical potential. This
corresponds to applying the Feshbach projection formalism to an Hilbert space
containing both particle and hole states.

 The present results are a first attempt at computing scattering processes
using the many-body Green's functions and required to introduce a certain
amount of phenomenological corrections.
 However, it is shown that predictions for the scattering of nucleons
can be obtained working also in the particle-hole space.
 The present results also give insight into the developments that will
be needed to pursue reliable microscopic calculations of the optical potential.
 We feel that the overall quality of the
results can be comparable to other methods applicable at
low energies~\cite{BayeDes,SMECO16}
when the role of the missing ingredients, such as short-range correlations, 
is included.
Thus SCGF could be considered as a valid candidate for the study of selected
reactions at astrophysical energies.

\vspace{1.in}

\acknowledgments
It is a pleasure to acknowledge several useful discussions with
J.-M.~Sparenberg and W.~H.~Dickhoff. We also thank Prof. Dickhoff
for providing his results form Ref.~\cite{DMP}.
We thank the Institute for Nuclear Theory at the University of Washington
for its hospitality during the initial writing of the manuscript.
This work was supported by the Natural
Sciences and Engineering Research Council of Canada (NSERC).



\end{document}